\documentclass[prb,twocolumn]{revtex4}
\usepackage{graphicx}
\usepackage{amsmath}

\newcommand{\up}{\uparrow}
\newcommand{\dn}{\downarrow}
\newcommand{\ket}[1]{|{#1}\rangle}
\newcommand{\bra}[1]{\langle {#1}|}
\newcommand{\av}[1]{\langle {#1} \rangle}
\newcommand{\etal}{{\it et al.} }

\begin{document}
\title{Impurity Potential Renormalization by Strong Electron Correlation}
\author{Noboru Fukushima}
\affiliation{
Institute of Physics, Academia Sinica, NanKang, Taipei 11529, Taiwan}

\author{Chung-Pin Chou}
\affiliation{
Institute of Physics, Academia Sinica, NanKang, Taipei 11529, Taiwan}
\author{Ting Kuo Lee}
\affiliation{
Institute of Physics, Academia Sinica, NanKang, Taipei 11529, Taiwan}

\begin{abstract}
Renormalization of non-magnetic impurity potential by strong electron
correlation is investigated in detail.
We adopt the $t$-$t'$-$t''$-$J$ model and consider mainly a
$\delta$-function impurity potential.
The variational Monte Carlo method shows that impurity potential
scattering matrix elements between Gutzwiller-projected quasi-particle
excited states are as strongly renormalized as the hopping terms.
Such renormalization is also seen by the Bogoliubov-de Gennes equation
with an impurity, where the strong correlation is treated by a
Gutzwiller mean-field theory with local renormalization factors and local
chemical potentials.  Namely, the $\delta$-function potential is
effectively weakened and broadened.
We emphasize the importance of including the local chemical potential,
which is paid little attention to in the literature, by physical
consideration of the doping dependence of a local hole density.
We also investigate effect of smooth impurity potential variation;
the strong correlation yields anticorrelation between the gap energy and
the coherence peak height simultaneously with large gap distribution,
which is consistent with the experiments.  \end{abstract}

\maketitle

\section{Introduction}

Anderson's theorem tells us that the $s$-wave superconductivity is
insensitive to small potential scattering \cite{Anderson59}. On the
other hand, the $d$-wave superconductivity has zero superconducting
gap in the nodal direction, and thus may be sensitive to disorder.
However, experimental observation of the high-temperature
superconductivity, which many people are nowadays convinced has
$d$-wave symmetry, seems robust against disorder
\cite{McElroyPRL05,PasupathySci08,ZhouPRL04,ShenSci05}.
For example, the high-temperature superconductors seem to exhibit more
conventional behavior at higher hole doping rates where the systems are
supposed to be more disordered.
Furthermore, the local density of states measured by the STM
\cite{Fang06,Kohsaka07} show clear V-shape at low energy that indicates the
$d$-wave nodes are not much influenced by disorder.
Theoretically, it is proposed that this protection of V-shape is due to
strong Coulomb repulsion between electrons
\cite{Anderson00,Garg08,FukushimaSNS}.  Hence, detailed studies of
effects of strong correlation for impurity scattering are necessary.

In correlated systems, the model parameters are effectively
renormalized.  For example, a hopping term is renormalized by
a factor smaller than unity because hopping is more difficult in the
presence of the double occupancy prohibition. On the other hand, an
exchange term is renormalized by a factor larger than unity because
each site is more often singly occupied. Then, the question is: how
are impurity terms renormalized?
In our previous paper\cite{FukushimaSNS}, we have presented preliminary
results of the impurity renormalization.  That is, local modulations of
the hopping and the exchange term tend to be enhanced by the local
renormalization factors, and impurity potential tends to be screened by
the effective local chemical potentials originated from minimization of
the total energy.
In this paper, we investigate in more detail how impurity potential is
renormalized, using the variational Monte Carlo (VMC) method and the
Gutzwiller-projected Bogoliubov-de Gennes (BdG)
equation\cite{QHWang06,CLi06,Fukushima08}, then discuss agreement and
disagreement with the experiments.

An extensive amount of literature has been devoted for the impact of
impurities on the normal and superconducting state of the cuprates.
A detailed review dedicated to this subject was recently given by
Alloul \etal \cite{Alloul09}
We do not repeat the whole review here, but the theoretical side may be
summarized as follows: Suppose electrons in host metal interact
with each other by the onsite repulsive Hubbard $U$ terms, and let us
put a {\it non-magnetic} impurity in it. Then, very strong impurity potential
(unitary scatterer), such as a cavity in otherwise uniform systems,
tends to induce local magnetic moments near the impurity if $U$ is
large enough, whereas weak potential (Born scatterer) does not.

However, the local moment formation in the superconducting state seems
relatively controversial.
The moments appear according to the theory based on the mean-field
decoupling of the $U$ term, e.g., by Chen and Ting \cite{YChen2004},
and Harter \etal \cite{Harter2007}.
Although it can be a good approximation for small $U$, yet
antiferromagnetic correlation is probably underestimated especially at
large $U$ because the superexchange process is not taken into account
explicitly.
Considering that the local moments typically appear at sufficiently
large $U$, it is critical to know in what range of $U$ the theory is valid.
An interesting contrast is in the theory by Tsuchiura \etal \cite{Tsuchiura01}
based on the one-site removed $t$-$t'$-$J$ model, i.e., an effective
$U\rightarrow\infty$ model.  They adopted two different Gutzwiller
approximations that lead to two different BdG equations.  One of them
takes away the double occupancy prohibition in return for the {\it
uniform} renormalization of model parameters.  This calculation results
in the local moment formation.  It also indicates that $J$ may cause the
appearance of the local moment even if $U=0$ because this effective
model is a ``$U=0$ but finite $J$'' system.
In the other BdG equation of them, each local model parameter is dressed
with an extended Gutzwiller renormalization factor that depends on the
position.  This calculation in contrast results in the absence of the local
moments because electrons tend to avoid the impurity and the
antiferromagnetism locally collapses.
However, these local renormalization factors are, without local
derivation, speculated from the previously derived formula for the {\it
uniform} system \cite{Ogata03}, and may need to be verified in the
future studies.
Tsuchiura's work was criticized by Wang and Lee \cite{ZWang02}, who
applied an inhomogeneous slave-boson mean-field theory to essentially
the same model.  The result shows the local magnetic moments in the
underdoped region.
In the slave-boson mean-field theory, the double occupancy prohibition
is relaxed by the saddle point approximation and only its average is
satisfied, which probably reduces the influence of the spin-spin interaction
effectively.  To compensate it, Wang and Lee added a phenomenological
residual spin-spin interaction term, and the result depends on its
magnitude.
In addition, Gabay \etal\cite{Gabay08} recently obtained similar results.
Liang and Lee\cite{Liang02} applied the VMC and also concluded
that local moments appear in the underdoped regime.

The strong scatterers introduced above are modeled on
in-plane impurities which substitute for Cu in the CuO$_2$ planes.
Besides such unitary scatterers, all cuprate materials are doped by
out-of-plane ions that mostly occupy random positions in the crystal
lattice or interstitial positions.
Such intrinsic impurities may be weak, but can be poorly screened by
electrons around them because the cuprates are quasi--two-dimensional
metal.  In addition, these weak potentials are not expected to induce
local magnetism.
Influence of these ``Born scattering potentials'' on the local density
of states was studied by, e.g., Wang and Lee,
\cite{QHWangDHLee03}, and Nunner \etal \cite{Nunner05}
In fact, however, the effect of electron correlation on them seems hardly
discussed in the literature, with the exception of Garg
\etal\cite{Garg08}, Fukushima \etal\cite{FukushimaSNS},
and Andersen and Hirschfeld \cite{Andersen08}.
That is a subject we would like to address in this paper.

After we define our model in Sec.~\ref{sec:model}, the renormalization of
impurity-potential matrix elements is shown by the VMC calculation in
Sec.~\ref{sec:VMC}.  Then, the BdG equation based on the Gutzwiller
approximation with local renormalization factors
\cite{QHWang06,CLi06,Fukushima08} are solved in Sec.~\ref{sec:BdG},
where we emphasize the importance of including local chemical potentials
through the comparison with the method by Garg \etal\cite{Garg08}
Note that these local chemical potentials are introduced for
minimizing the total energy and are different from those used for the
inhomogeneous slave-boson mean-field
theory\cite{ZWang02,Gabay08,Ziegler98} that are the Lagrange multipliers
to enforce the average no-double-occupancy condition.
Renormalization of smoothly varying impurity potential is also presented
to show anticorrelation between the gap energy and the coherence peak
height compatible with large gap distribution,
which is consistent with the experiments \cite{Fang06, McElroySci05}.

\section{Model}
\label{sec:model}

We use the $t$-$t'$-$t''$-$J$ model with an impurity term, namely,
\begin{gather}
H \equiv H_{tt't''} + H_{J} + H_{\rm imp},
\\
 H_{tt't''} \equiv P\left(- \sum_{i,j,\sigma} t_{ij}  c_{i\sigma}^\dagger
           c_{j\sigma} \right)P
\\
 H_{J} \equiv J \sum_{\langle i,j\rangle}
\left( {\bf S}_i\cdot {\bf S}_j -\frac14 \hat{n}_i \hat{n}_j \right),
\end{gather}
where $c_{i\sigma}^\dagger$ ($c_{i\sigma}$) is the creation
(annihilation) operator of site $i$ and spin $\sigma$, and
$\hat{n}_i \equiv \sum_\sigma c_{i\sigma}^\dagger c_{i\sigma}$.  As the
hopping, we take $t_{ij}=t, t', t'',$ for the nearest, second, and third
neighbors, respectively, and otherwise zero. The summation in the $J$
term is taken over every nearest-neighbor pair.  The Gutzwiller
projection operator $P$ prohibits electron double occupancy at every
site.

In this paper, we focus on the renormalization of a single non-magnetic
$\delta$-function impurity potential located at $i=0$,
\begin{equation}
 H_{\rm imp} = V_0\sum_\sigma c_{0\sigma}^\dagger c_{0\sigma}
= \frac{V_0}{N_L} \sum_{k,k'\sigma} c_{k\sigma}^\dagger c_{k'\sigma},
\end{equation}
except for Sec.~\ref{sec:smooth}, where we briefly discuss
the effect of smoothly varying impurity potential.
Here, $N_L$ is the number of sites.

\section{Variational Monte Carlo calculation for matrix element renormalization}
\label{sec:VMC}

Here we calculate matrix elements of the impurity potential with
respect to the uniform Gutzwiller-projected quasi-particle states of
$d$-wave superconductors using the VMC method.

Let us start from a uniform system without impurities.  We assume that
the ground state is well approximated by a Gutzwiller-projected $d$-wave
superconducting state,
\begin{equation}
|{\rm GS} \rangle\propto P \left[\sum_p \frac{v_p}{u_p} \,
c_{p\up}^\dagger c_{-p\dn}^\dagger \right]^{\frac{N_e}{2}} |0\rangle
,
\end{equation}
where
\begin{gather}
u_k \equiv \sqrt{ \frac{1}{2} \left( 1 + \frac{\xi_k}{E_k}  \right)} ,
\quad
v_k \equiv \frac{\Delta_k}{|\Delta_k|}
\sqrt{ \frac{1}{2} \left( 1 - \frac{\xi_k}{E_k}  \right)} ,
\nonumber\\
E_k \equiv \sqrt{ \xi_k^2 + \Delta_k^2 } , \quad
\Delta_k \equiv \Delta_{\rm v}\,(\cos k_x -\cos k_y ) , \nonumber \\
\xi_k \equiv -2 t_{\rm v} (\cos k_x +\cos k_y)
 -4 t_{\rm v}' \cos k_x \cos k_y  \nonumber \\
 -2 t_{\rm v}''(\cos 2k_x +\cos 2k_y)
 -\mu_{\rm v} . \nonumber
\end{gather}
$N_{e}$ is the total number of electrons. The variational parameters
$\Delta_{\rm v}$, $t_{\rm v}'$, $t_{\rm v}''$, $\mu_{\rm v}$ are
optimized so as to minimize the total energy.
We also assume that the excited states are well represented by the
projected quasi-hole wave function\cite{CPChouJMMM07},
\begin{gather}
|k\sigma\rangle\propto P \, c_{k\sigma}^\dagger \left[\sum_p
\frac{v_p}{u_p} c_{p\up}^\dagger c_{-p\dn}^\dagger
\right]^{\frac{N_e}{2}-1}|0\rangle.
\end{gather}
Then, we are able to calculate the matrix elements and spectral
weights using the excited quasi-hole wave function with
the ground state parameters.

By switching on the impurity potential, these excited states should
be mixed by the matrix elements,
\begin{equation}
V_{k,k'}\equiv\langle
k\uparrow|c^\dagger_{k\uparrow}c_{k'\uparrow}|k'\uparrow\rangle+\langle
-k'\uparrow|c^\dagger_{k\downarrow}c_{k'\downarrow}|-k\uparrow\rangle.
\label{eq.defVkk}
\end{equation}
We carry out VMC calculation for $V_{k,k'}$,
and show that its renormalization by the strong correlation
is similar to that of the total spectral weight,
\begin{equation}
Z_k \equiv \Big|\bra{k\sigma} c_{k\sigma}^\dagger \ket{\rm GS}
\Big|^2 + \Big |\bra{k\sigma} c_{-k -\sigma} \ket{\rm GS} \Big|^2 ,
\end{equation}
which is known to be strongly renormalized \cite{CPChou06}.
It is worthy to be noted that the BCS
theory has predicted the matrix elements of the impurity potential and
the total spectral weights are $V_{k,k'}^{\rm BCS} = u_k u_{k'} -
v_k v_{k'}$, and $Z_k^{\rm BCS} = 1$, respectively. On the other
hand, the Gutzwiller approximation (GA) yields renormalization,
\begin{equation} Z_k^{\rm GA} = g_t \equiv \frac{2x}{1+x}, \end{equation}
where $g_t$ is the Gutzwiller renormalization factor for the hopping term.
However, according to the conventional GA, $V_{k,k'}$ is not
renormalized because it originally comes from a particle number
operator; on the contrary, a GA generalized for inhomogeneous systems
yields renormalization as will be discussed in the next section.

\begin{figure}[h]
\begin{center}
\includegraphics[width=7cm]{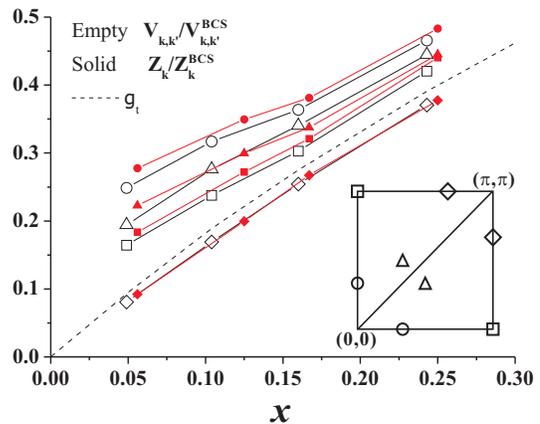}
\end{center}
\caption{(Color online) Comparison of $ V_{k,k'}/V_{k,k'}^{\rm BCS}
$ and $ Z_{k}/Z_{k}^{\rm BCS} $ as functions of the hole
concentration $x$ in the case of $(t',t'',J)/t=(-0.3,0.2,0.3)$
optimized at each $x$.  Each symbol represents transfer between the
$k$-points of the same symbols in the inset. \label{fig:VMC:VandZ}}
\end{figure}

\begin{figure}[h]
\begin{center}
\includegraphics[width=7cm]{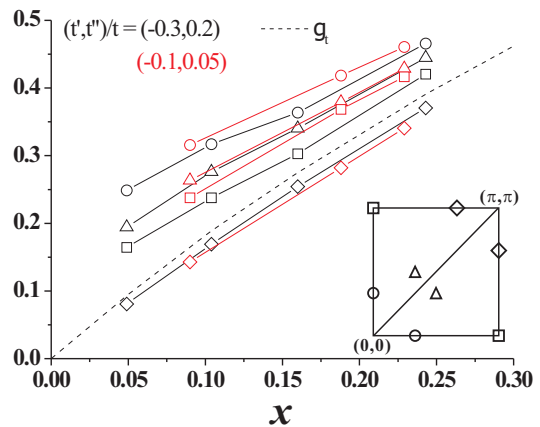}
\end{center}
\caption{(Color online) Comparison of $V_{k,k'}/V_{k,k'}^{\rm BCS}$
of different bare parameters.  Each symbol represents transfer
between the $k$-points of the same symbols in the inset.
\label{fig:VMC:Vrenorm}}
\end{figure}

We plot $ V_{k,k'}/V_{k,k'}^{\rm BCS} $ and $ Z_{k}/Z_{k}^{\rm BCS}
$ in Fig.~\ref{fig:VMC:VandZ} as functions of the hole
concentration. Suppose the impurity potential is not too large.
Then, the matrix elements $V_{k,k'}$ perturb the system only if the
excitation energies of $\ket{k\sigma}$ and $\ket{k'\sigma}$ are
close. Therefore, here we plot $V_{k,k'}$ connecting two symmetric
reciprocal lattice points indicated in the inset of
Fig.~\ref{fig:VMC:VandZ}; each pair of symbols in the inset refers
to $k$ and $k'$, which corresponds to the same symbols of $V_{k,k'}$
and $Z_k=Z_{k'}$ in the plot. The variational parameters are
optimized for each hole concentration.
Note that $V_{k,k'}$ is renormalized as strongly as $Z_k$,
and its renormalization factor is quite close to $g_t$.
Furthermore, Fig.~\ref{fig:VMC:Vrenorm} compares $ V_{k,k'} $ of
different bare parameters in the Hamiltonian.  It suggests that the
renormalization is insensitive to parameters.


These results may be understood as follows.
Let us look at the Fourier transform of the $\delta$-function
impurity potential [Eq.~(\ref{eq.defVkk})]. The sum of its diagonal
terms $(k=k')$ just slightly shifts the chemical potential. What
about the off-diagonal terms $(k \neq k')$? If $k$ were a site
index, they would be renormalized as $g_t = 2x/(1+x)$ according
to the Gutzwiller approximation, which is smaller than unity and is
going to zero as $x\rightarrow 0$ because it is more difficult to
hop in the presence of the double occupancy prohibition. Even in $k$
space, if electrons are densely packed in the lattice, it must be
similarly difficult to hop from $k$ to a different $k'$. Thus, the
impurity potential should be renormalized by a factor similar to
$g_t$ as we expected.

\section{Gutzwiller-projected Bogoliubov-de Gennes equation}
\label{sec:BdG}

\subsection{Renormalization by effective local chemical potentials}
\label{sec:localchem}

We solve a BdG equation derived using the Gutzwiller approximation
with local renormalization factors\cite{QHWang06,CLi06,Fukushima08} for
non-magnetic systems. By requiring minimization of the total energy,
the BdG Hamiltonian naturally contains effective local chemical
potentials originating from the derivative of the local renormalization
factors with respect to local particle densities.  In the following,
we show that the impurity potential is renormalized by those local
chemical potentials.

Let us assume that a good variational ground state can be
represented in the form of $P'|\psi\rangle$, where $|\psi\rangle$
represents a wave function obtained later by solving a BdG equation.
The operator $P'$ contains a fugacity operator to control the
particle number as well as the original Gutzwiller projection $P$.
In the following, we use notation,
\begin{equation}
\av{ \hat{O} }_0 \equiv \frac{\langle \psi| \hat{O} |\psi\rangle}
     {\langle \psi|\psi\rangle},
     \quad
\av{ \hat{O} } \equiv \frac{\langle \psi|P' \hat{O} P'|\psi\rangle}
     {\langle \psi|P'P'|\psi\rangle},
\end{equation}
for an arbitrary operator $\hat{O}$.
The built-in fugacities allow us to require conservation of the
local electron densities, namely,
\begin{equation}
\av{ \hat{n}_i } = \av{ \hat{n}_i }_0 \equiv n_i
. \label{eq:niconserve}
\end{equation}
Then, the Gutzwiller approximation yields
\begin{align}
& \av{ c_{i\sigma}^\dagger c_{j\sigma} }
 \simeq
g^t_{ij} \av{ c_{i\sigma}^\dagger c_{j\sigma} }_0, & g^{t}_{ij}
\equiv \sqrt{\frac{2x_i}{1+x_i}\cdot\frac{2x_j}{1+x_j}},
 \label{eq:gutzt}
\\
& \av{ {\bf S}_i\cdot {\bf S}_j }
 \simeq g^s_{ij} \langle {\bf
S}_i\cdot {\bf S}_j \rangle_0 , & g^{s}_{ij} \equiv
\frac{2}{1+x_i}\cdot\frac{2}{1+x_j}, \label{eq:gutzs}
\end{align}
where $x_i\equiv 1-n_i $.

We consider non-magnetic systems
where $\chi_{ij} \equiv \langle c^\dagger_{i\up}c_{j\up} \rangle_0
= \langle c^\dagger_{i\dn}c_{j\dn} \rangle_0$ and
$\Delta_{ij} \equiv \langle c_{j\downarrow}c_{i\uparrow} \rangle_0
$ are real numbers, and $\Delta_{ij}=\Delta_{ji}$.
Then, the total energy $\av{H -\mu\sum_i \hat{n}_i}$ can be
calculated by the GA as
\begin{gather}
E_{\rm GA} =
-4 \sum_{(i,j)} g^{t}_{ij} t_{ij} \chi_{ij}
-\sum_{\langle i,j \rangle}
\frac{J}{4}\big[
2(3 g^s_{ij}-1) \chi_{ij}^2
 \nonumber \\
+ 2(3 g^s_{ij}+1) \Delta_{ij}^2
+ n_{i}n_{j} \big]
-\mu \sum_i n_i +V_0 n_0,
\end{gather}
where the summation of the kinetic-energy term is taken over every
$(i,j)$ pair.
Using $ \hat{\chi}_{ij}\equiv \sum_{\sigma} (c_{i\sigma}^\dagger
c_{j\sigma} + c_{j\sigma}^\dagger c_{i\sigma}) /4 $ and
$\hat{\Delta}_{ij}\equiv( c_{i\uparrow}^\dagger c_{j \downarrow}^\dagger
+ c_{j\uparrow}^\dagger c_{i \downarrow}^\dagger + c_{j\dn} c_{i
\up} + c_{i\dn} c_{j \up} )/4 $, the
extremum condition of $E_{\rm GA}$ leads to a BdG equation
\cite{QHWang06,Fukushima08} represented by the mean-field
Hamiltonian $ H_{\rm BdG} = \sum_{(ij)}\hat{\chi}_{ij} dE_{\rm
GA}/d\chi_{ij} + \sum_{\langle ij \rangle}\hat{\Delta}_{ij} dE_{\rm
GA}/d\Delta_{ij} + \sum_i \hat{n}_{i} \, dE_{\rm GA}/dn_{i} $, namely,
\begin{gather}
 H_{\rm BdG} =
- \sum_{ij\sigma}  g^{t}_{ij} t_{ij} c_{i\sigma}^\dagger c_{j\sigma}
- \sum_{\langle ij \rangle} J (3g^{s}_{ij}-1)  \chi_{ij} \hat{\chi}_{ij}
\nonumber\\
- \sum_{\langle ij \rangle} J (3g^{s}_{ij}+1)  \Delta_{ij} \hat{\Delta}_{ij}
-\sum_{i} (\mu+\mu_i)  \hat{n}_i +V_0 \hat{n}_0
.
\label{eq:BdG}
\end{gather}
Note that,
in contrast to the conventional BdG equation,
it contains the effective local chemical potential
\begin{eqnarray}
 \mu_i &\equiv &  \frac{d E_{\rm GA}}{dx_i} - \mu
= -\sum_{j} 4 \frac{dg^{t}_{ij}}{dx_i} t_{ij} \chi_{ij}
\nonumber \\ &&
  -\sum_{j \rm{(n.n.)}} \left[
\frac32 \frac{dg^{s}_{ij}}{dx_i}
  J(\chi_{ij}^2 + \Delta_{ij}^2) -\frac{J}{4} n_j
\right],
\end{eqnarray}
where the summation of the $J$ term is taken over the nearest
neighbors of site $i$.
By diagonalizing the BdG Hamiltonian, we obtain
$H_{\rm BdG}=\sum_{n=1}^{N_L} E_n (\gamma_{1n}^\dag\gamma_{1n} +
\gamma_{2n}^\dag\gamma_{2n}) + {\rm const.}$, with $E_n\ge 0$ and
\begin{equation}
\left(\begin{array}{c}
\gamma_{1n} \\
\gamma_{2n}^\dagger \\
\end{array}\right)=\sum_i
\left(\begin{array}{cc}
u^{n}_{i} & v^{n}_{i} \\
- v^{n}_{i} & u^{n}_{i} \\
\end{array}\right)\left(\begin{array}{c}
c_{i\uparrow} \\
c^{\dag}_{i\downarrow} \\
\end{array}\right).
\end{equation}
Then, $|\psi\rangle= \prod_n \gamma_{1n} \gamma_{2n} |0\rangle$,
$n_i=2\sum_n (v^{n}_{i})^2$, $\chi_{ij}= \sum_n  v^{n}_{i} v^{n}_{j}
$, and $ \Delta_{ij}= \sum_n  - u^{n}_{i} v^{n}_{j} $.

The inclusion of the local chemical potential $\mu_i$ makes it harder
to optimize the local mean-fields, and simple iteration does not
converge very well.  Strategies to look for the minimum of the total
energy $E$ seem to work slightly better.
We have solved the self-consistent equation for the systems of
24$\times$24 sites with the periodic boundary condition.  We set
$t'=-0.3t$ and $t''= 0.2t$.  Then, $J$ and $\mu$ are determined using
the uniform system without the impurity so that
$x$ is the desired hole concentration as well as
 $ J (3g^{s}_{ij}+1) \Delta_{ij} = 0.3t$.
These values are fixed
in solving the equation for the impurity systems, i.e., we neglect
$O(1/N_{L})$ shift of $\mu$ caused by the inclusion of the
impurity.

\begin{figure}[h]
 \begin{center}
\includegraphics[width=8cm]{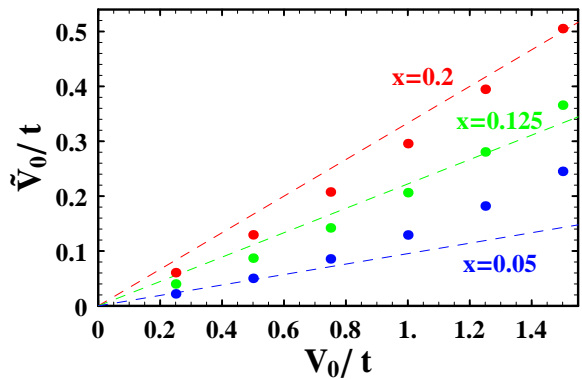}
 \end{center}
\caption{(Color online) The dots represent the renormalized impurity
 potential at the impurity site as a function of the bare impurity
 potential with the hole concentrations $x=0.05$ (blue), $0.125$ (green),
 and $0.2$ (red).  The broken lines are $g_t V_0$.  \label{fig:VvsV} }
\end{figure}

\begin{figure}[h]
 \begin{center}
\includegraphics[width=8cm]{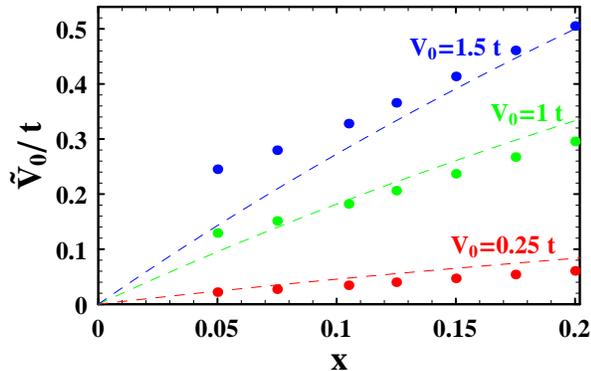}
 \end{center}
\caption{(Color online) The dots represent the renormalized impurity
 potential at the impurity site as a function of the hole concentration
 with the bare impurity potentials $V_0=0.25t$ (red), $1t$ (green), and
 $1.5t$ (blue).  The broken lines are $g_t V_0$.  \label{fig:Vvsx} }
\end{figure}

According to Eq.~(\ref{eq:niconserve}), the expectation value of
$\hat{n}_i$ is by definition not renormalized by any ``$g$'' factor
as the hopping and the exchange term. However, as one can see in
the BdG Hamiltonian in Eq.~(\ref{eq:BdG}), the impurity potential can
be compensated by $\mu_i$. Therefore, we define a renormalized
impurity potential by including difference of $\mu_{i}$, namely,
\begin{equation}
 \tilde{V_i}=V_0 \delta_{i0}-\left(\mu_{i}-\mu_{\infty} \right)
.
\end{equation}
Here, $\mu_{\infty}$ is $\mu_{i\rightarrow\infty}$ and
approximately equal to $\mu_{i}$ of the system without the impurity,
which is nonzero \cite{FCZhang88}.  In the uniform system,
however, one usually redefines $\mu + \mu_{\infty}$ as the chemical
potential, and $\mu_{\infty}$ does not explicitly appear in the
calculation \cite{FCZhang88}.
Figures \ref{fig:VvsV} and \ref{fig:Vvsx} show the calculated
renormalized impurity potential at the impurity site for various
values of the bare potential $V_0$ and the hole concentration $x\ge
0.05$.
We have also tried the systems with $x = 0.025$,
but were unable to reach the energy minimum possibly because
there are a couple of meta-stable states.


\begin{figure}[h]
 \begin{center}
\includegraphics[width=8cm]{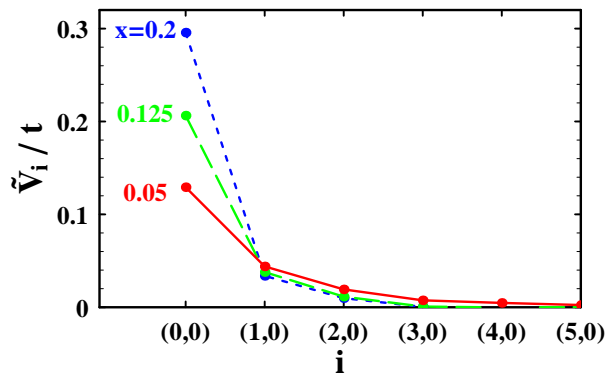}
 \end{center}
\caption{(Color online) The position dependence of
the renormalized impurity potential for $V_0=1t$.
\label{fig:Vvsi}
}
\end{figure}

Note that $\tilde{V}_0$ is strongly suppressed and is quite close to $g_t
V_0$, where $g_t$ is the Gutzwiller renormalization factor of the uniform system.  These
results agree with our VMC results in the previous section.
Here, $\tilde{V}_0$ deviates upward from $g_t V_0$ when $V_0$ is large
near the half filling.  This is possibly related to the position
dependence of $\tilde{V}_i$.
Originally, the impurity potential is nonzero only at the impurity
site.  However, after solving the BdG equation, the renormalized
impurity potential distributes more broadly as shown in
Fig.~\ref{fig:Vvsi}.
Namely, as $V_0$ becomes larger and $x$ becomes smaller,
the renormalization effect reduces at the impurity site
whereas it broadens toward the neighboring sites.

This broadening can be understood as follows: Basically, energy loss by
the impurity potential reduces electron occupation at the impurity site.
However, the hole prefers to move around to gain the kinetic energy.
Therefore, to minimize the total energy, the $\delta$-function impurity
potential is broadened by $\mu_i$.
More explicitly, the contribution to $\mu_i$ from the kinetic energy
contains a factor
\begin{equation}
 \frac{dg^{t}_{ij}}{dx_i}=
\sqrt{\frac1{2x_i (1+x_i)^{3}}}  \, \sqrt{\frac{2x_j}{1+x_j}}
,
\end{equation}
which behaves as $\sqrt{x_j/x_i }$ near the half filling.
Suppose the local hole densities behave as
$x_i\sim x^{\delta_i}$,  $x_j\sim x^{\delta_j}$
with some exponents $\delta_i$, $\delta_j$.
Then, $dg^{t}_{ij}/dx_i \sim x^{(\delta_j-\delta_i)/2}$
as $x\rightarrow0$.
If $\delta_i\neq\delta_j$, then $\mu_i$ or $\mu_j$ diverges
and the self-consistent condition is not satisfied.
Therefore, $\delta_i=\delta_j$,
which tends to make the hole distribution more uniform.

Such extension of the impurity potential is also reported in the context
of the unitary impurity potential by Poilblanc \etal\cite{Poilblanc94}
and in the weak-coupling context by Ziegler \etal\cite{Ziegler96} In
addition, Bulut \cite{Bulut00} and Ohashi \cite{Ohashi01} reported that
the agreement between experimental data and the random-phase approximation
is improved by adding phenomenological extended range potential.

At half filling, the non-magnetic impurity potential should not affect
the ground state because each site has to be occupied by one
electron in any case.  However, they do affect the ground-state
energy. In the words of the BdG equation,
the impurity potential is renormalized by $\mu_i$ in $H_{\rm BdG}$,
and not in $E_{\rm GA}$.
This is different from the well-known renormalization of the hopping and the
exchange term described by $g^t_{ij}$ and $g^s_{ij}$; these renormalization
factors influence both the ground state and the ground-state energy.

\subsection{Local density of states}

The projected quasi-particle states $P'\gamma_{\sigma n}^\dag|\psi\rangle$
are approximately orthogonal to each other\cite{Fukushima08}. We
regard them as excited states and calculate the density of states (DOS).
Then, the local DOS (LDOS) is represented by
\begin{gather}
N(r,\omega) = g^t_{rr}\sum_n
\left[
  |u_r^n|^2 \delta(\omega-E_n)
+ |v_r^n|^2 \delta(\omega+E_n)
\right]
.
\end{gather}

To obtain dense spectra, we use a supercell composed of
24$\times$24 sites whose origin has the impurity, and this supercell is
repeated so as to construct a superlattice of 10$\times$10 supercells
with the periodic boundary condition\cite{Tsuchiura00}. Then, the
Hamiltonian can be block-diagonalized by the Fourier transform with
respect to the supercell indices, and the calculation of expectation values
is reduced to an average over many quasi-twisted boundary conditions of
the 24$\times$24 site system.
This supercell method is useful to obtain dense energy levels
although it seems to overestimate correlation functions between very
distant sites if the supercell size is too small and the number of
supercells is too large.
Except for this supercell boundary condition, the other conditions of the
calculation are the same as those in Sec.~\ref{sec:localchem}.
Since spectra in finite systems are discrete, we replace each
$\delta$-function by the Gaussian distribution with the standard
deviation $\delta E=0.02t$ to obtain continuous DOS.

\begin{figure}[h]
 \begin{center}
\includegraphics[width=8.5cm]{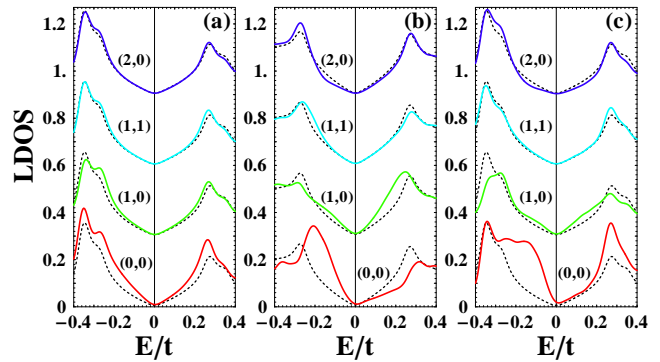}
 \end{center}
\caption{(Color online) The LDOS
for one impurity system with $V=1t$ and $x=1/8$,
at the impurity site and its
nearest, second, and third neighbor,
calculated by three different BdG equations:
(a) with $g^t_{ij}$, $g^s_{ij}$, $\mu_i$, (b) without correlation
($g^t_{ij}=g^s_{ij}=1$, $\mu_i=0$), (c) without the local chemical
potential\cite{Garg08} (with $g^t_{ij}$, $g^s_{ij}$, but $\mu_i=0$).
The dotted lines are the DOS in the uniform system ($V_0=0$) as a
reference.
\label{fig:LDOS} }
\end{figure}

Figure \ref{fig:LDOS}(a) shows the calculated LDOS at the impurity
site as well as its nearest, second, and third neighbors.  The DOS of
the uniform system ($V_0=0$) is also plotted by dotted lines as a
reference.
Here, we have chosen $V_0=1t$ as the impurity potential, which is
of the same order as the renormalized band width ($8 g_t t$).
Nevertheless, it is well screened by $\mu_i$ and the LDOS is not very
site-dependent in agreement with results in Sec.~\ref{sec:localchem}.
At the impurity site, the hole density is larger, and thus $g^t_{rr}$ is
large. As a result the LDOS is also larger than those of the other
sites.
The peak at $E\sim-0.37t$ is the van Hove peak; the band
renormalization by $g_t$ makes it closer to the Fermi surface.
In fact, another van Hove peak appears in the positive bias at
$E\sim0.37t$. It is much weaker than the negative-bias van Hove
peak, and only appears as a small shoulder of the the coherence peak
in the resolution of Fig.~\ref{fig:LDOS}(a). Some small portion of
the original electron band around the van Hove peak is unoccupied in
the mean-field superconducting ground state due to the electron-hole
mixture.  Even though that portion is small, the singular DOS
enhances it to yield the positive-bias van Hove peak.

For comparison, we also show in Fig.~\ref{fig:LDOS}(b) the results of
the system without the strong correlation, i.e, let us set
$g^t_{ij}=g^s_{ij}=1$ and $\mu_i=0$ in Eq.~(\ref{eq:BdG}) and solve the
BdG equation.  $J$ and $\mu$ are determined in the same way as in the
strongly correlated case, i.e., $4 J \Delta_{ij}=0.3t$ for the uniform
system.
Here, we use supercells of $36\times36$ sites to obtain dense spectra;
if the same system size is used, energy level spacings are larger than
those in the correlated systems due to the wider band width.
The impurity potential is also $V_0=1t$.  Note that it is this time much
smaller than the band width $8 t$.  Nevertheless, comparing
Figs.~\ref{fig:LDOS}(a) and \ref{fig:LDOS}(b),
it is clear that the system without
strong correlation is much more disordered than the one with the
correlation.

From Fig.~\ref{fig:LDOS}(b), it is clear that the $\delta$-function
potential makes the LDOS asymmetric.\cite{QHWangDHLee03,Nunner05} That
is, some weight of the LDOS tends to move to the right at the impurity
site, and to the left at the nearest neighbor site.  The shift is large
at high energy, and small at low energy.
This asymmetry appears because the $\delta$-function potential lifts the
degeneracy of two quasi-particles: (i) a linear combination between an
electron state at site 1 and a hole state at site 2, or (ii)
a linear combination between an electron state at 2 and a hole state at 1.
We discuss it more in detail using a two-site system in Appendix
\ref{sec:twosite}.
With the strong correlation in Fig.~\ref{fig:LDOS}(a), however, the
asymmetry is less pronounced and the LDOS seems more symmetric.
Most likely, the broadening of the impurity potential by $\mu_i$ results
in weakening of the asymmetry.

\subsection{Comparison with other strongly-correlated BdG equations}

\subsubsection{Importance of including the local chemical potential $\mu_i$}

Garg et al.\ used a similar BdG equation\cite{Garg08}.
Although the definition of the local renormalization factors is the same,
they do not take into account the effective local chemical potential
$\mu_i$ that minimize the total energy.
They have reported that the spatially averaged LDOS
shows protected V-shapes at low energy, and stated
that the correlated systems are less disordered.
For the testing purpose, we have also solved their BdG equation and show
the resultant LDOS in Fig.~\ref{fig:LDOS}(c).  The parameters are the
same as Fig.~\ref{fig:LDOS}(a), and the uniform limits of these two BdG
equations are identical.
We have found that the LDOS {\it before} the spatial average is
actually quite disordered, especially at the impurity site, and that
the V-shaped LDOS appears only after the average.  Therefore, it
seems difficult to conclude that the LDOS in Fig.~\ref{fig:LDOS}(c)
is less disordered than the LDOS in Fig.~\ref{fig:LDOS} (b), in our
opinion.

\begin{figure}[h]
 \begin{center}
\includegraphics[width=8cm]{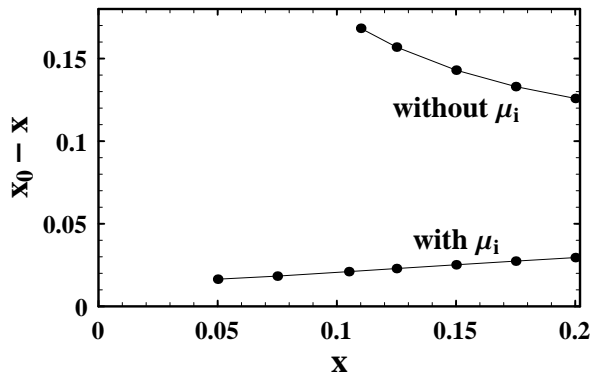}
 \end{center}
\caption{
The relative hole density $x_0-x$ at the impurity site as a function of
the bulk hole density $x$ solved for $V_0=0.25t$ by the two different
 BdG equations; without $\mu_i$\cite{Garg08} and with $\mu_i$.
\label{fig:xvsx}
 }
\end{figure}

A more serious problem of their method appears in the local hole density
as a function of the bulk hole density shown in Fig.~\ref{fig:xvsx}.
As already mentioned above, the non-magnetic impurity potential at half
filling should not affect the ground state because each site has to be
occupied by one electron in any case.  Hence one can naturally imagine
that the local hole density $x_0$ at the impurity site approaches the
bulk hole density $x$ as $x\rightarrow 0$.
However, if $\mu_i$ is not taken into account, $x_0-x$ increases as $x$
decreases.
Therefore, it is questionable if the method by Garg {\it et al.}\ correctly
captured the properties of the strongly correlated systems.

Capello \etal\cite{Capello08} solved the BdG equation with extended
Gutzwiller renormalization factors, but without the local chemical potential.  Although
we have not duplicated their results, we speculate that they have
similar problems as Garg {\it et al.}\ because of the lack of the local
chemical potential to minimize the total energy.

\subsubsection{Large $U$ instead of the Gutzwiller projection}

Andersen and Hirschfeld\cite{Andersen08} and some references therein
used the BdG equation without the Gutzwiller renormalization factors ($g_t
=g_s=1$).  They took into account the electron correlation by
the Hubbard $U$ term of the mean-field level.
In that case, the screening effect similar to the one described in this
paper can be obtained because $-U n_{i\bar\sigma}$ plays a role of our
local chemical potential $\mu_i$.
Since the impurity potential reduces electron occupation, the Coulomb
energy loss is smaller there.  Namely, summing up the potential and
Coulomb energy loss, the total loss at the impurity site become less
prominent by $U$.
Note that, however, the strength of the screening depends on how one
chooses $U$.
A problem is that the mean-field decoupling of the $U$ term may
underestimate the exchange energy and overestimate the kinetic energy
especially near the half filling; the ground state would be strongly
influenced by them.

\subsection{Smoothly varying impurity potential}
\label{sec:smooth}

As has been shown above, short-range impurity potential is screened by $\mu_i$.
Accordingly, what remains must be the smooth variation in the
potential.
Here, we demonstrate the influence of the strong electron correlation
on it using the Coulomb potential.

Instead of $H_{\rm imp}$, we consider the Coulomb potential from
randomly located off-plane impurities, namely,
\begin{gather}
H_{\rm smooth} \equiv
\sum_{r \sigma} V_r c_{r\sigma}^\dagger c_{r\sigma},
\\
V_r\equiv \sum_{\ell=1}^{N_{\rm imp}}
\frac{V_{\rm C}}{\sqrt{(r-r_\ell)^2+d^2}}
,
\label{eq:Coulombpot}
\end{gather}
where $N_{\rm imp}$ is the number of impurities, $r_\ell$ is the
position of $\ell$-th impurity projected on the $xy$ plane, and $d$
is the off-plane distance.
Here, we take $V_{\rm C}=1t$ and $d=1$  and adjust $\mu$ to
satisfy $x=1/8$ simultaneously with the self-consistency condition.
In addition, one supercell has 20$\times$20 sites with $N_{\rm imp}=12$
impurities, and the same impurity configuration is repeated so as to
construct a superlattice of 10$\times$10 supercells.  For simplicity, in
determining the Coulomb potential in Eq.~(\ref{eq:Coulombpot}), we use
only one of the cells, i.e., the system of 20$\times$20 sites with the
periodic boundary condition.

\begin{figure}[h]
 \begin{center}
\includegraphics[width=8cm]{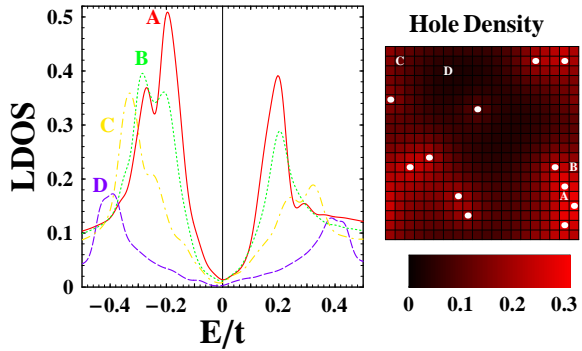}
 \end{center}
\caption{(Color online) (Left) The LDOS at 4 sites A, B, C and D
 indicated in the right figure, for the system with smoothly varying
 potential with $x=1/8$, $V_{\rm C}=1t$, $d=1$, $N_{\rm imp}=12$, and
 supercell size 20$\times$20 sites, solved by the Gutzwiller-projected
 BdG equation. (Right) The hole density distribution in a supercell. The
 white dots are positions of the impurities.  \label{fig:smooth} }
\end{figure}

\begin{figure}[h]
 \begin{center}
\includegraphics[width=8cm]{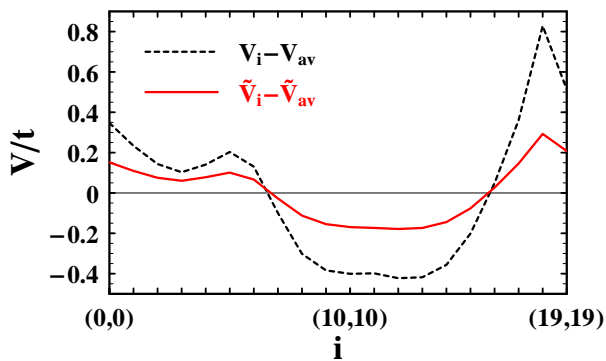}
 \end{center}
\caption{(Color online) Spatial dependence of the bare impurity
potential $V_i-N_L^{-1}\sum_i V_i$ and the renormalized impurity
potential $\tilde{V}_i-N_L^{-1}\sum_i \tilde{V}_i$ along a diagonal line
(from the left bottom to the right top in the hole density plot of
Fig.~\ref{fig:smooth}). \label{fig:Vvsismooth} }
\end{figure}

Figure \ref{fig:smooth} shows the LDOS at 4 sites
A, B, C, and D, each of which has a different hole concentration.  The
LDOS in hole-rich regions (e.g., A) has high coherence peaks with low gap
energy. In contrast, hole-poor regions (e.g., D) has low coherence peaks
with high gap energy.
Figure \ref{fig:Vvsismooth} compares the spatial dependence of the bare
impurity potential $V_r-N_L^{-1}\sum_r V_r$ and the renormalized
impurity potential $\tilde{V}_r-N_L^{-1}\sum_r \tilde{V}_r$.  Here, we
have subtracted the spatial average of the potential.  It is clear that
the renormalized impurity potential has much smaller and smoother
variation.

Nunner \etal\cite{Nunner05} pointed out for the conventional BdG
equation with smoothly varying potential that the LDOS at site $r$ is
similar to the DOS in the uniform system with shifted chemical potential
$\tilde{\mu}=\mu-V_r$.
This argument can be also applied to our system.  However, the
Gutzwiller renormalization factors make a difference in the relation among the hole
density $x_r$, the gap energy, and the height of the coherence peaks.
In the systems without the Gutzwiller projection,
$\Delta_{ij}$ is determined by the DOS near the Fermi level. Namely,
$\Delta_{ij}$ takes maximum when the chemical potential is at the van
Hove singularity.  Accordingly, when $\Delta_{ij}$ is large, the gap
energy and the peak height are also large.
In contrast, with the Gutzwiller renormalization factors, as $x\rightarrow 0$, (i) the
band shrinks by $g^t_{ij}$ and thus the DOS near the Fermi level
increases, and (ii) the pairing interaction is enhanced by
$g^s_{ij}$.  Because of (i) and (ii), $\Delta_{ij}$ and the gap
energy monotonically increases as $x\rightarrow 0$.
Furthermore, the LDOS contains an extra factor $g^t_{ii}$, and thus the
peak height decreases as $x$ decreases.

Note that this anticorrelation between the gap energy and the peak
height as well as the large gap distribution is consistent with the
experiments \cite{Fang06,McElroySci05}.  However, according to our calculation,
the large spatial variation in the LDOS accompanies large variation in the
hole density, which has yet to be verified by the experiments.

\section{Summary and discussion}
\label{sec:summary}
In this paper, we have investigated
the renormalization of the impurity potential by the strong correlation.
The VMC calculation has shown that impurity potential scattering
matrix elements between Gutzwiller-projected quasi-particle excited
states are as strongly renormalized as the hopping terms.
It may be understood by the Fourier transform of the $\delta$-function
impurity potential having a form of the hopping term in $k$ space.
The (real-space) hopping term is known to be renormalized by a factor
less than unity because it is more difficult to hop in the presence of
the double occupancy prohibition.
Even in $k$ space, if electrons are densely packed in the lattice, it
must be similarly difficult to hop from $k$ to $k'\neq k$, then the
impurity potential and the hopping term should be renormalized
similarly.

Such reduction in the impurity potential is also seen by the BdG
equation with local Gutzwiller renormalization factors.
Near the half filling of the strongly correlated systems, the influence of
the non-magnetic impurity potential on the ground state is small because
each site has to be occupied by almost one electron in any case.
However, the impurity potential {\it does} affect the ground-state energy.
Such properties appear by taking into account effective local chemical
potential, which is paid little attention to in the literature.
In addition, the local chemical potential effectively broadens the
impurity potential because holes prefer to move around to gain the
kinetic energy.
Effect of smoothly varying impurity potential has been
briefly discussed.  It shows large gap distribution.  If the
Gutzwiller renormalization factors are taken into account, the gap energy and the
peak height are anticorrelated.  These properties are consistent
with the experiments \cite{Fang06, McElroySci05}.

In fair comparison, there are also some disagreements with the
experiments.
According to our results, short-range non-magnetic potential variations
are reduced, thus the system is more uniform, and accordingly the
$d$-wave superconductivity can be robust.  However, in the experiments,
the system seems quite disordered but still the $d$-wave is robust.
Such short-range disorder may be introduced by spatial modulation of
$t_{ij}$ or $J$, which can be enhanced by the strong correlation
\cite{FukushimaSNS}, or by magnetic impurities, or by the
electron-lattice interaction \cite{CPChou09}.

In addition, although the smooth impurity potential variation yields
anticorrelation between the gap energy and the peak height, it does
not show the almost spatially independent V-shaped LDOS at low energy
seen often in the experiments, which can be explained instead in the
case of the rapid potential variation\cite{CPChou08}.
In our previous paper\cite{CPChou08}, we have discussed the LDOS of
stripe states, where $\Delta_{ij}$ contains two components; one is
spatially uniform and the other is oscillating, typically with wave
number $q=\pi/4$ or $\pi/2$.
Then, the V-shaped gap is determined by the uniform component, and the
oscillating component influences it little.  As a result, the linear
slope of V-shape is
robust (it does not have much spatial dependence).  This is rather
counterintuitive because one may think that the local gap could be
determined by $\Delta_{ij}$.  However, the superconducting gap is not
determined by such local properties, but by ``coherence'', i.e., spatial
dependence of $\Delta_{ij}$.
Let us recall that, in the case of the zero-momentum pairing as the
conventional BCS theory, the spin-up electron band couples with the
spin-down hole band (upside-down spin-down electron band); these
bands intersect at the Fermi level, and a gap opens by a nonzero
superconducting order parameter.
On the other hand, the oscillating components contain only pairing with
nonzero center-of-mass momentum.  Then, the spin-up electron band
couples with the $q$-shifted (or multiples-of-$q$ shifted) spin-down
hole band.  The point is, these bands typically intersect not at the Fermi level
except for limited points.  In such cases, the V-shape of the LDOS is
not affected a lot.

Similarly, oscillations of the variational parameters other than
$\Delta_{ij}$ with wave number $q$ mix ``the bare band'' and ``the bands
shifted by multiples of $q$''.  Such terms change the band structure
especially near band intersections.
However, this change in the band structure is not related to the
superconductivity, at least in the mean-field approximation.
The superconducting DOS depends on where one puts the chemical potential,
but general properties near the V-shaped LDOS
is determined by $\Delta_{ij}$.

The smooth impurity potential variation considered in Sec.~\ref{sec:smooth}
is similar to the stripe states with $q\simeq 0$.  Since $q$ is
small, the oscillating components of $\Delta_{ij}$ has effect
similar to the uniform component $q=0$.
Namely, it affects every state at the Fermi level.
However, the gap has now spatial dependence because $q$ is not
completely zero.  In this case, the local gap is determined by
$\Delta_{ij}$;
the LDOS at sites $i$ is similar to the DOS in the uniform system
with $\tilde{\Delta}^x=-\tilde{\Delta}^y = \Delta_{i,i+\hat{x}}$,
$\tilde{\chi}=\chi_{ij}$, and $\tilde{\mu}=\mu-V_i$.
Therefore, there is nothing like the ``uniform component'' in the
argument of the stripe state, and thus
the linear slope of V-shape in
the LDOS is not robust anymore.
This issue will be addressed in the future.

\acknowledgments
The authors thank C.-M.\ Ho for stimulating discussions.
This work was supported by the National Science Council in Taiwan with
Grant no.95-2112-M-001-061-MY3.
Part of the numerical calculation was performed in the IBM Cluster 1350,
the Formosa II Cluster and the Triton Cluster in the National Center for
High-performance Computing in Taiwan,
and the PC Farm II of Academia Sinica Computing Center in Taiwan.

\appendix

\section{ LDOS asymmetry caused by $\delta$-function potential}
\label{sec:twosite}

As shown in Fig.~\ref{fig:LDOS}, the $\delta$-function potential
causes asymmetry in the LDOS near the impurity site.  Such asymmetry may
be important because some STM experiments analyze data by taking the
ratio between intensities of positive and negative bias
\cite{Kohsaka07}: the symmetric part of the LDOS is canceled and its
asymmetric part remains.
To explain the origin of the asymmetry, let us diagonalize the BdG
Hamiltonian of a simple two-site problem analytically in the
following. We know that it is not realistic to apply the mean-field
approximation to two-site problems.  However, it provides us some
insight into numerical solutions in the bulk systems.

Suppose the potentials at sites 1 and 2 are $V$ ($>0$) and $-V$,
respectively.
By putting the states in the order of electrons
${1\uparrow}$, ${2\uparrow}$, and holes ${1\downarrow}$,
${2\downarrow}$, the BdG Hamiltonian matrix is written as
\begin{equation}
 H_{\rm 2 site} =
\begin{pmatrix}
V & -t & 0 & \Delta \\
-t & -V & \Delta & 0 \\
0 & \Delta & -V & t \\
\Delta & 0 & t & V
\end{pmatrix}
.
\end{equation}
Let us start from a simple case of $t=0$ and assume $V<\Delta$.
Then, $\Delta$ mixes ${1\uparrow}$ electron and ${2\downarrow}$ hole
with the equal weights.  We call the linear combination of them with
positive and negative energy the quasi-particle A$^+$ and A$^-$,
respectively.
Similarly, B$^\pm$ denotes the linear combination of ${2\uparrow}$
electron and ${1\downarrow}$ hole with positive/negative energy.

If $V=0$,  A and B are degenerate. Finite but small $V$ causes energy
difference between A and B, namely, $E_{{\rm A}^\pm}=\pm \Delta +
V$, $E_{{\rm B}^\pm}=\pm \Delta - V$.
In the ground state, A$^-$ and B$^-$ are occupied, but A$^+$ and B$^+$ are
unoccupied.
First we focus on 1$\uparrow$.  Then, the electron addition spectrum is
at $E=E_{{\rm A}^+}$, the removal is at $E=E_{{\rm A}^-}$. Namely, the
finite V just shifts the whole spectra by $V$, and the local spectra are
not symmetric around $E=0$.
On the other hand, the spectra of site 2 shift by $-V$.
As a result, the spectra summed over sites 1 and 2 are
still symmetric.  As for 1$\downarrow$, the addition is at $-E_{{\rm
B}^-}=E_{{\rm A}^+}$, and the removal is at $-E_{{\rm B}^+}=E_{{\rm
A}^-}$ (these negative signs originate from the treatment of
1$\downarrow$ as a hole), which are identical to the spectra of
1$\uparrow$ as expected.

Next, let us consider $t\ne 0$.  Then, $t$ mixes A$^+$ and B$^-$,
and the eigenstates are linear combinations of them whose
eigenenergies are $ \pm [t^2+\left(\Delta + V \right)^2]^{1/2} $.
The eigenenergies of superposition of A$^-$ and B$^+$ are $ \pm
[t^2+\left(\Delta - V \right)^2]^{1/2} $.
As $t$ increases, the LDOS asymmetry becomes weaker, but it never
disappears.

Back to the bulk systems,
maybe we can physically explain the asymmetry as follows.
Since $J$ term causes pairing, a Cooper pair is formed more or
less between nearest neighbors (site 1,2)
when a snapshot is taken.
This Cooper pair is a resonance of ``states where sites 1 and 2 are
occupied by electrons'' and ``states where both are occupied by
holes''.
Destruction of the pair leaves a quasi-particle.  There are two
possibilities for it; (i) an ``electron at site 1 and hole at site 2'', and
(ii) an ``electron at 2 and hole at 1''.
The $\delta$-function potential lifts the degeneracy of these
quasi-particles.  That results in the asymmetry in the local spectra
although the bulk spectra are symmetric as in the two-site problem above.
In the bulk system, the spectra are continuous, and the spectral shift by
the impurity potential is larger at high energy than at low energy because
(i) the near-nodal quasi-particles at low energy are less influenced by
the impurity potential because there are not many states to mix with,
and (ii) the shift is caused by $\Delta$ and it is smaller at lower
energy.


\begin{thebibliography}{99}

\bibitem{Anderson59} P. W. Anderson, J. Phys. Chem. Solids {\bf 11} 26 (1959).
\bibitem{PasupathySci08} A. N. Pasupathy, {\it et al.}, Science {\bf 320}, 196 (2008).
\bibitem{McElroyPRL05} K. McElroy, {\it et al.}, Phys. Rev. Lett. {\bf 94}, 197005 (2005).
\bibitem{ZhouPRL04} X. J. Zhou, {\it et al.}, Phys. Rev. Lett. {\bf 92}, 187001 (2004).
\bibitem{ShenSci05} K. M. Shen, {\it et al.}, Science {\bf 307}, 901 (2005).

\bibitem{Fang06}
A. C. Fang, L. Capriotti, D. J. Scalapino, S. A. Kivelson, N. Kaneko,
M. Greven, and A. Kapitulnik,
Phys. Rev. Lett. {\bf 96} 017007 (2006).

\bibitem{Kohsaka07} Y. Kohsaka {\it et al.}, Science {\bf 315} 1380
	(2007).

\bibitem{Anderson00} P. W. Anderson, Science {\bf 288} 480 (2000).
\bibitem{Garg08} A. Garg, M. Randeria, N. Trivedi,
Nature Phys. {\bf 4}, 762 (2008).
\bibitem{FukushimaSNS}
N. Fukushima, C.-P. Chou, T. K. Lee,
J. Phys. Chem. Solids {\bf 69}, 3046 (2008).

\bibitem{QHWang06}
Q.-H. Wang, Z.D. Wang, Y. Chen and F.C. Zhang,
Phys. Rev. B {\bf 73}, 092507 (2006).

\bibitem{CLi06}
 C. Li, S. Zhou, and Z. Wang,
 Phys. Rev. B {\bf 73} 060501(R) (2006).

\bibitem{Fukushima08}
N. Fukushima, Phys. Rev. B {\bf 78}, 115105 (2008).


\bibitem{Alloul09}
 H. Alloul, J. Bobroff, M. Gabay, P. J. Hirschfeld,
 Rev. Mod. Phys. {\bf 81}, 45 (2009).

\bibitem{YChen2004}
 Y. Chen and C. S. Ting,
 Phys. Rev. Lett. {\bf 92}, 077203 (2004).

\bibitem{Harter2007}
J. W. Harter, B. M. Andersen, J. Bobroff, M. Gabay, and P. J. Hirschfeld,
Phys. Rev. B {\bf 75}, 054520 (2007).

\bibitem{Tsuchiura01}
H. Tsuchiura, Y. Tanaka, M. Ogata, and S. Kashiwaya,
Phys. Rev. B {\bf 64}, 140501(R) (2001).

\bibitem{Ogata03}
 M. Ogata and A. Himeda,
 J. Phys. Soc. Jpn. {\bf 72}, 374 (2003).

\bibitem{ZWang02}
 Z. Wang and P. A. Lee,
 Phys. Rev. Lett. {\bf 89}, 217002 (2002).

\bibitem{Gabay08}
M. Gabay, E. Semel, P. J. Hirschfeld, and W. Chen,
 Phys. Rev. B {\bf 77}, 165110 (2008).

\bibitem{Liang02}
 S.-D. Liang and T. K. Lee,
 Phys. Rev. B {\bf 65}, 214529 (2002).

\bibitem{QHWangDHLee03} Q.-H. Wang and D.-H. Lee,
Phys. Rev. B {\bf 67}, 020511(R) (2003).

\bibitem{Nunner05}
T.S. Nunner, B.M. Andersen, A. Melikyan, and P.J. Hirschfeld,
Phys. Rev. Lett. {\bf 95}, 177003 (2005).

\bibitem{Andersen08}
B. M. Andersen and P. J. Hirschfeld,
Phys. Rev. Lett. {\bf 100}, 257003 (2008).


\bibitem{Ziegler98}
 W. Ziegler, H. Endres, and W. Hanke,
 Phys. Rev. B {\bf 58}, 4362 (1998).

\bibitem{McElroySci05}
K. McElroy {\it et al.}, Science {\bf 309}, 1048 (2005).


\bibitem{CPChouJMMM07} C.-P. Chou, T. K. Lee, and C.-M. Ho, J. Magn. Magn. Mater. {\bf 310}, 474 (2007).
\bibitem{CPChou06} C.-P. Chou, T. K. Lee, C.-M. Ho, Phys. Rev. B {\bf 74}, 092503 (2006).

\bibitem{FCZhang88}
 F. C. Zhang, C. Gros, T. M. Rice, H. Shiba,
 Supercond. Sci. Technol. {\bf 1}, 36 (1988).

\bibitem{Poilblanc94}
D. Poilblanc, D. J. Scalapino, and W. Hanke,
Phys. Rev. B {\bf 50}, 13020 (1994).

\bibitem{Ziegler96}
W. Ziegler, D. Poilblanc, R. Preuss, W. Hanke, D. J. Scalapino,
Phys. Rev. B {\bf 53}, 8704 (1996).


\bibitem{Bulut00}
N. Bulut,
Phys. Rev. B {\bf 61}, 9051 (2000).

\bibitem{Ohashi01}
Y. Ohashi,
J. Phys. Soc. Jpn. {\bf 70}, 2054 (2001).


\bibitem{Tsuchiura00} H. Tsuchiura, Y. Tanaka, M. Ogata, and S. Kashiwaya,
Phys. Rev. Lett. {\bf 84}, 3165 (2000).



\bibitem{Capello08} M. Capello, M. Raczkowski, and D. Poilblanc, Phys. Rev. B {\bf 77}, 224502 (2008).

\bibitem{CPChou09} C.-P. Chou and T. K. Lee, in preparation.

\bibitem{CPChou08} C.-P. Chou, N. Fukushima, and T. K. Lee,
Phys. Rev. B {\bf 78}, 134530 (2008).

\end{thebibliography}
\end{document}